# 5G Network Architecture and Configuration Choices to Support Teleoperated Driving at Scale

M.C. Lucas-Estañ[1], B. Coll-Perales[1], M. I. Khan[2], J. Gozalvez[1], S. S. Avedisov[2], O. Altintas[2], M. Sepulcre[1]
[1]Uwicore laboratory, Universidad Miguel Hernandez de Elche, Elche (Alicante), Spain.
[2]InfoTech Labs, Toyota Motor North America R&D, Mountain View, CA, U.S.A.
{m.lucas, bcoll, j.gozalvez, msepulcre}@umh.es; {mohammad.irfan.khan, sergei.avedisov, onur.altintas}@toyota.com

*Abstract*— **Teleoperated driving (ToD) enables the remote driving or control of vehicles. For this purpose, vehicles must transmit video feeds to the ToD control center so that the remote operator is fully aware of the driving conditions and can safely control the vehicle. 5G (and beyond) networks are fundamental for the deployment of ToD as they can provide the low latency, reliable and broadband connection necessary to connect the vehicle and ToD control center. However, it is unclear whether common 5G network architectures and configurations are well-suited to support the simultaneous teleoperation of multiple vehicles with demanding uplink bandwidth, as current networks are mainly configured to support mobile broadband services. This paper demonstrates that MEC or edge-based 5G networks are better suited to support and scale the ToD service than centralized networks, and quantifies the bandwidth required to simultaneously teleoperate multiple vehicles under various 5G network architectures and configurations, including different duplexing modes and TDD frame structures. Finally, the study shows that the configuration of the control channels can help mitigate the impact that the processing time of the video feeds has on the capacity to support and scale the ToD service.**

## I. INTRODUCTION

Teleoperated Driving (ToD) offers the capability to remotely drive or control vehicles, assisting them in navigating challenging driving conditions. With ToD, an operator can take over control of vehicles from a ToD control center (remote control), and either drive the vehicle directly or provide a driving path for the vehicle to navigate a challenging situation (remote guidance). ToD requires vehicles to transmit video feeds and perception data to the ToD control center so that the operator has a comprehensive and accurate understanding of the driving conditions and environment. The resulting driving commands must be transmitted with low latency and jitter for a safe control of the vehicle.

5G (and beyond) networks play a critical role in ToD as they can provide the low latency, reliable and broadband connection necessary to connect the vehicle and the ToD control center. The potential of 5G to support ToD has been demonstrated in pilots developed, for example, in EU projects 5GCroCo [1] and 5GMobix [2]. Pilots are generally conducted under controlled conditions with a limited number of vehicles, and the 5G network is typically configured following the Global System for Mobile Communications Association (GSMA) recommendations [3] and common commercial 5G deployments. For example, [1] utilizes a centralized 5G network where the ToD control center is remotely connected through an Internet connection. The network operates on 40 MHz with a Time Division Duplexing (TDD) asymmetric frame structure reserving 4 slots for the downlink (DL) and 1 slot for the uplink (UL). Centralized networks increase the latency, while asymmetric TDD frame structures limit the capacity to upload broadband video feeds from the vehicle to the ToD control center [4].

Several studies have analyzed the feasibility of mobile networks to support ToD. For example, [5] conducted a large measurement campaign to evaluate whether LTE networks could support ToD. Similarly, [6] and [7] report network measurements over a commercial 5G network that supports a teleoperated vehicle, and [8] measures the latency experienced when transmitting a video stream via a 5G Non-Public Network for the remote control of a monorail vehicle. However, existing studies focus on a single ToD vehicle, and it remains unclear whether common 5G network deployments and configurations can scale the ToD service, and for instance, support the target density of 10 ToD vehicles per km$^2$ established by 5G Automotive Association (5GAA) in [9]. This study advances the state-of-the-art by analyzing the capacity of different 5G network architectures and configurations to simultaneously support multiple ToD vehicles. We demonstrate that Multi-Access Edge Computing (MEC) or edge-based 5G network architectures are better suited to support ToD than centralized networks, which are constrained by the Internet connection to the ToD control center. We quantify the bandwidth required to scale the ToD service and support the 5GAA target density under centralized and MEC-based 5G network architectures. We also show that the bandwidth required to scale the ToD service is strongly influenced by the duplexing mode and the configuration of the TDD frame structure, particularly by the number of slots reserved for downlink and uplink transmissions and the time between consecutive uplink slots. Finally, we demonstrate that the configuration of the 5G control channels can mitigate the impact of the processing time of the video feeds transmitted from the vehicles to the ToD control center on the capacity to support and scale the ToD service.

## II. TOD SERVICE REQUIREMENTS

5GAA defines in [9] four possible ToD use cases for remote control or guidance, and this study focuses on the use case where a remote operator directly controls a vehicle from a ToD control center as it is the use case with the most stringent network demands. The vehicle collects perception data (e.g., video feeds) and transmits this data to the ToD control center. The remote operator sends driving commands to the vehicle. ToD with direct control requires the transmission of perception data from the vehicle to the control center through a base station (or gNB) in the uplink (UL) direction. It also requires the transmission of driving commands from the control center to the vehicle in the downlink (DL) direction with low latency and high reliability.

5GAA defines the ToD service requirements for UL and DL traffic in [9], and these requirements are summarized in

This work was supported in part by MCIN/AEI/10.13039/501100011033 (grant PID2020-115576RB-I00), by the "European Union NextGenerationEU/PRTR" (TED2021-130436B-I00), and by Generalitat Valenciana, and UMH's Vicerrectorado de Investigación grants.

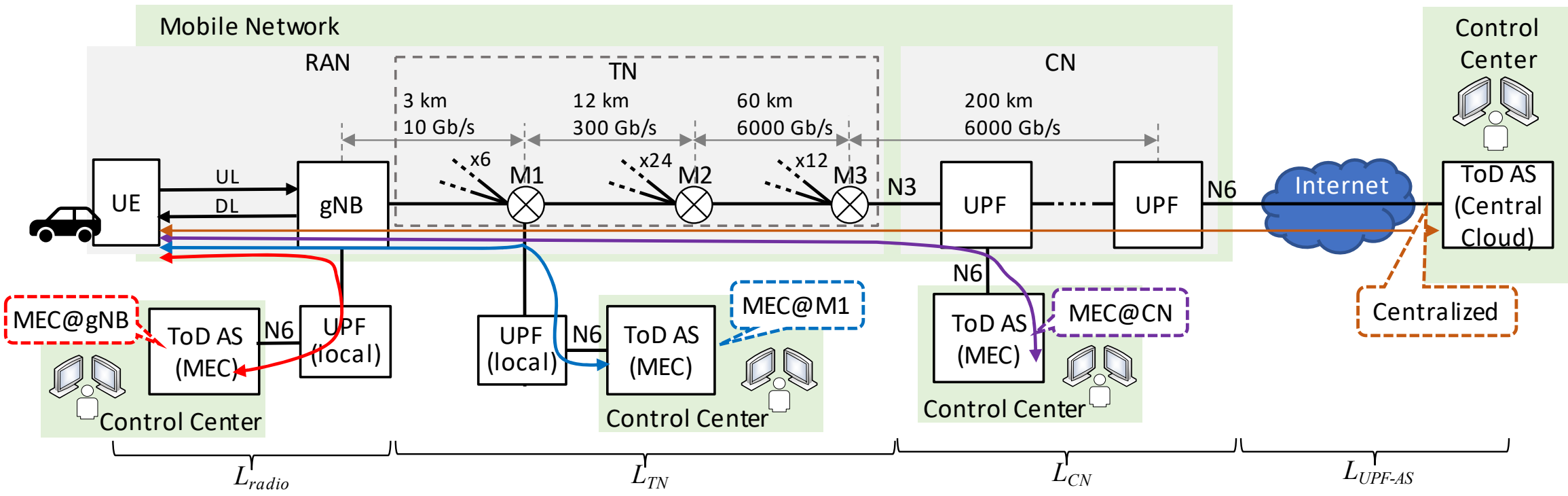

**Fig. 1**. 5G network architectures for ToD service.

Table I for the ToD use case analyzed in this study. Vehicles are expected to transmit video feeds at 8 Mbps from four cameras to the ToD control center (UL). Meanwhile, the ToD control center should transmit 8 Kb command messages to the vehicle every 20 ms (400 Kbps) in the DL. The latency requirement is defined in [10] as *measurements of time from the occurrence of the event in scenario application zone to the beginning of the resulting actuation*. Reliability is defined as the percentage of packets successfully delivered within the latency requirement. Table I shows that the DL traffic has more stringent latency and reliability requirements than UL traffic due to the critical nature of the driving commands. Additionally, 5GAA establishes that the ToD service should support a target density of ToD vehicles of 10 veh/km$^2$.

Following [10], the ToD latency requirement includes the processing delays $Lp$ and the communications latency from transmitter to receiver ($L_{UL}$ and $L_{DL}$, for UL and DL, respectively). On the uplink, the processing delay includes the time to capture ($Tc$) and encode ($Te$) one video frame at the vehicle, as well as the time to decode ($Td$) and render ($Tr$) a video frame at the ToD control center. In this context, $Lp$ represents the video processing latency, which is the sum of $Tc$, $Te$, $Td$ and $Tr$. [10] estimates that $Lp$ equals 57 ms when cameras generate 60 frames per second (fps), and approximately 90 ms when the frame rate reduces to 30 fps. Another study [11] models each component of $Lp$ using normal distributions, resulting in a total average value of $Lp$ equal to 40 ms. We should note that $Lp$ can significantly vary depending on the implementation of the encoders employed for video processing. Ensuring low video processing latencies is important as it sets the limit of the maximum UL communication latency to meet the UL latency requirement. On the downlink, the processing delay includes the car command execution delay, which is 10 ms based on [10]. Table II lists the latency variables previously defined.

TABLE I. ToD SERVICE LEVEL REQUIREMENTS [9]

| Data Rate | Latency | Reliability | Target density |
|---|---|---|---|
| UL: 32 Mbps<br>DL: 400 Kbps | UL: 100 ms<br>DL: 20 ms | UL: 99%<br>DL: 99.999% | 10 veh/km$^2$ |

TABLE II. LATENCY VARIABLES

| Variable | Definition |
|---|---|
| $Lp$ | Processing delay $Lp = Tc + Te + Td + Tr$ |
| $L_{UL}$, $L_{DL}$ | Communications latency from transmitter to receiver in UL and DL respectively |
| $Tc$, $Te$ | Time to capture ($Tc$) and encode ($Te$) a video frame at the vehicle |
| $Td$, $Tr$ | Time to decode ($Td$) and render ($Tr$) a video frame at the ToD control center |
| $L_{radio}$ | Latency experienced at the radio access network |
| $L_{TN}$ | Latency experienced at the transport network |
| $L_{CN}$ | Latency experienced at the core network |
| $L_{UPF\text{-}AS}$ | Latency experienced in the connection between the UPF in the CN and the V2X application server in the ToD control center |

## III. 5G COMMUNICATION LATENCY

We analyze the capability of 5G networks to support the ToD service by utilizing and extending the 5G communication latency models presented in [12] and [13], which combine simulation, queuing theory, and statistical models derived from empirical studies to estimate the various latency components included in the model. These models enable us to estimate the communication latency experienced during the transmission of video feeds from the vehicle to the control center ($L_{UL}$) and the transmission of driving commands from the control center to the ToD vehicle ($L_{DL}$).

The communication latency is estimated for centralized and MEC-based 5G network architectures illustrated in Fig. 1. We consider the network topology recommended by the International Telecommunication Union (ITU) [14] with a hierarchical Transport Network (TN) in the Radio Access Network (RAN) connecting the core network (CN). The TN comprises three nodes (M1, M2 and M3), which multiplex traffic from 6 gNBs, 24 M1 and 12 M2 nodes, respectively. In the centralized architecture, the ToD control center is deployed at a central server with an Internet connection. The ToD traffic traverses through the gNB, TN, CN, and the Internet from/towards the vehicle towards/from the ToD control center. Alternatively, the ToD control center can be implemented within a MEC in the mobile network operators' domain, which may be located at the first User Plane Function (UPF) in the CN (MEC@CN in Fig. 1), at node M1 in the TN (MEC@M1), or connected to a gNB (MEC@gNB). The closer the ToD control center is to the vehicles, the lower the communication latency. However, the area controlled by one ToD control center will be smaller.

The 5G communication latency model accounts for the latency experienced at the radio, TN and CN ($L_{radio}$, $L_{TN}$ and $L_{CN}$, respectively), as well as the latency introduced by the connection between the UPF in the CN and the V2X application server (AS) in the ToD control center ($L_{UPF\text{-}AS}$), as illustrated in Fig. 1 and expressed in (1):

$$L_{UL/DL} = L_{radio} + L_{TN} + L_{CN} + L_{UPF\text{-}AS} \quad (1)$$

The defined latency variables are reported in Table II. The implementation of these latency components varies depending on the 5G network architecture, with the main difference lying in the connection to the application server (AS). In a centralized network architecture, $L_{UPF\text{-}AS}$ corresponds to the latency introduced by an Internet connection. Internet connections increase the latency (and its variability) compared to MEC-based network architectures, where the ToD control center is hosted in the mobile network operator's domain.

$L_{radio}$ is determined using the model presented in [12], which simulates the radio resource allocation process and packet transmission (using a customized implementation in Matlab) considering the influence of the 5G NR numerology, Sub-Carrier Spacing (SCS), Modulation and Coding Schemes (MCSs), and scheduling and retransmission mechanisms. The model was initially developed for Frequency Division Duplexing (FDD), and is extended in this study to estimate the radio latency for different TDD frames. Additionally, the model has been enhanced with packet segmentation capabilities to accommodate the transmission of large packets generated by video cameras in ToD vehicles.

$L_{TN}$ and $L_{CN}$ denote the propagation and transit delays across the TN and CN, and are estimated using [13]. The propagation delay measures the time packets need to traverse TN or CN links. The transit delay accounts for the time needed for receiving, processing, and transmitting packets at TN or CN nodes. We used queueing theory to calculate the transit delay, which depends on the number of nodes the packets traverse, network load, and the links' capacity.

$L_{UPF\text{-}AS}$ represents the latency introduced by the Internet connection to the ToD center for the centralized architecture. It is modeled using an empirical study [13] that characterizes the round-trip time observed between source-target Internet nodes within the same country. In the case of MEC-based architectures, the link between the UPF and application server (AS) is part of the CN, and hence $L_{UPF\text{-}AS}$ is already part of $L_{CN}$.

## IV. Evaluation Scenario

We consider a highway scenario with three lanes per direction and two densities of ToD vehicles (4 and 10 veh/km$^2$). 10 veh/ km$^2$ is the target density of ToD vehicles that 5G networks should support according to the 5GAA requirements defined in [9]. [9] establishes that each vehicle should transmit video feeds at 8 Mbps from four cameras. However, we demonstrated in [4] that it is not feasible to scale the ToD service with such bit rates in the case of a centralized 5G architecture. In this case, it is necessary to reduce the video bit rates to support the target density of ToD vehicles established by 5GAA. Therefore, this study considers the transmission of video feeds at 1.125 Mbps per camera, resulting in a total bitrate of 4.5 Mbps per vehicle. This video bitrate per vehicle was utilized in ToD trials in European projects such as [1]. We employ the model presented in [15] to generate realistic H.264 video frames from each vehicle's camera. The frames are generated periodically at 30 fps, and the model determines the size of the video frames using a Weibull distribution. Fig. 2 compares the cumulative distribution function of the size of the video frames generated with the H.264 video model against those obtained from videos captured in the CARLA simulator or available in the KITTI dataset from field measurements. The figure shows that the distribution generated with the model aligns closely with the distributions derived from KITTI and CARLA, thus validating the H.264 model.

The scenario considers a single 5G cell with a radius of 866 meters, and we analyze all four possible 5G network architectures depicted in Fig. 1. The distances and link capacities between TN nodes are shown in Fig. 1. We reserve a fraction of the links' capacities for ToD traffic. This fraction is calculated to prevent a queue backlog of V2X packets at the TN and CN nodes. We follow 3GPP guidelines [16] for configuring the radio access network as reported in Table III.

We analyze the latency with three TDD frame structures and FDD. The first two structures are recommended by GSMA in [3] and are commonly used in most 5G deployments. The first TDD frame structure is DDDSU and was used in [1]; we refer to this structure as TDD1 throughout the paper. D and U denote slots allocated for DL and UL transmissions, respectively. S is a special slot that can distribute the 14 Orthogonal Frequency-Division Multiplexing (OFDM) symbols between DL and UL transmissions, along with guard periods. Therefore, TDD1 allocates 3 consecutive slots for DL transmissions, followed by 1 special slot, and 1 slot for UL transmissions. The second TDD frame structure is DDDDDDDSUU and we refer to it as TDD2. TDD2 allocates 7 consecutive slots for DL transmissions, followed by 1 special slot, and 2 consecutive slots for UL transmissions. TDD1 and TDD2 allocate most of the symbols in S to DL transmissions, so we treat the S slot as a D slot in this study. The third TDD frame structure is DUDU (TDD3) and defines a more balanced allocation of slots for UL and DL. TDD3 aligns with 5G standards but its adoption remains limited as current 5G network deployments primarily focus on supporting broadband mobile services with higher demand for DL bandwidth. A more balanced allocation is also achieved with FDD, which equally distributes the bandwidth between UL and DL.

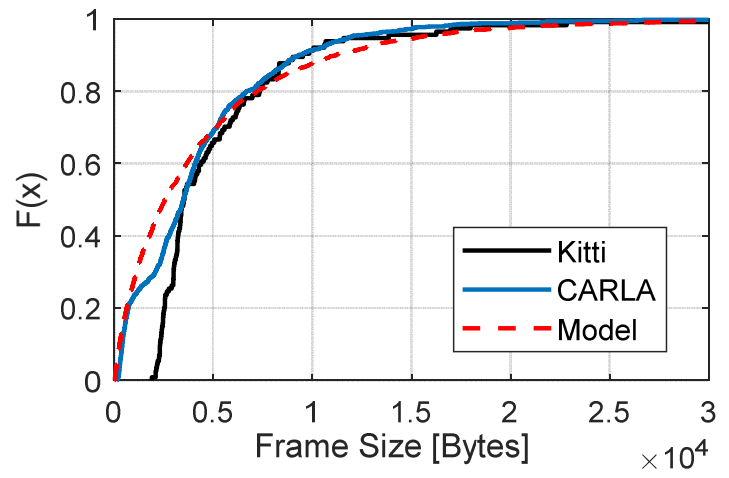

**Fig. 2**. Cumulative distribution function of the size of the video frames generated with the H.264 model and captured using CARLA and KITTI.

TABLE III. Radio Access Network Configuration Parameters

| Parameter/Mechanism | Value |
|---|---|
| Sub-Carrier Spacing (SCS) | 30 kHz |
| Multiple input multiple output (MIMO) layers | 2 |
| Full/single-slot transmission (tx) | Full-slot |
| Scheduling | UL: dynamic DL: Semi-Persistent |
| Modulation and Coding Schemes (MCS) table (see [17]) | UL: MCS Table 2, DL: MCS Table 3 |
| Target Block Error Rate (BLER) | UL: 0.01, DL: 0.00001 |
| Hybrid automatic repeat request (HARQ) max. number of retx | UL: 3, DL: 1 |
| Bandwidth (*BW*) | 30, 40, 50, 60, 70, 80, 90, 100 MHz |

## V. Impact of the 5G Network Architecture

This section analyzes the feasibility of supporting ToD under centralized and MEC-based 5G network architectures. To this end, we use the 5G communication latency model presented in Section III. Our primary focus is on the MEC@CN architecture, as it requires a significantly lower number of ToD control centers to support and scale the ToD service. Additionally, deploying the MEC connected to a gNB only reduces the communication latency by 1 ms compared to deploying it at the CN. We evaluate separately the capacity to support the ToD service for UL and DL since their ToD requirements significantly differ, and most 5G deployments asymmetrically allocate resources for UL and DL [1].

### A. Uplink communications

We first analyze the percentage of UL packets that fulfill the 5GAA requirements necessary to support the ToD service, as outlined in Table I. Specifically, 5GAA mandates that 99% of UL packets are correctly received at the control center in less than 100 ms. This analysis is performed considering the TDD1 frame structure and varying video processing delays *Lp* ranging from 40 ms to 90 ms. Fig. 3 shows the percentage of packets meeting the 100 ms latency requirement for different *Lp* values under vehicle densities of 4 and 10 veh/km$^2$ and variable 5G bandwidths. We should note that as the value of *Lp* increases, the maximum tolerable UL communications latency to meet the 100 ms requirement decreases. The figure shows that, except for low video processing delays (*Lp*), MEC-based 5G network architectures ensure a significantly higher percentage of packets meeting the ToD latency requirement compared to a centralized architecture. It is important to note that these differences increase with higher values of *Lp*, indicating that MEC-based architectures can accommodate larger video processing delays. For instance, in a centralized architecture, only 21.6% of packets meet the ToD latency requirement when *Lp* is 80 ms, the density is 4 veh/km$^2$, and the 5G bandwidth is 40 MHz. In contrast, this percentage increases to 97.8% for the MEC@CN network architecture. Additionally, it is worth noting that the MEC@CN network architecture requires significantly less bandwidth to achieve the same performance. For instance, while the centralized architecture needs 60 MHz when *Lp* is equal to 60 ms to meet the latency requirement for 99% of the packets with 4 veh/km$^2$, the MEC@CN network architecture can sustain the same performance with only 40 MHz.

MEC-based network architectures can better guarantee the ToD UL latency requirements due to their lower communication latency. Fig. 4 depicts the 99$^{th}$ percentile of the UL communication latency for the centralized and MEC-based architectures when the MEC is located at the CN, M1 node in the TN, or the gNB. The figure omits any value when the number of packets received is lower than 99%. Fig. 4 shows that the 99$^{th}$ percentile is approximately 20 ms higher for the centralized architecture than for the MEC-based architectures. This difference is attributed to the latency introduced by the Internet connection between the CN and the ToD control center. Additionally, Fig. 4 reveals that similar latencies are achieved across the MEC-based architectures. This is because all the MEC-based architectures experience the same latency at the radio link, which constitutes more than 93% of the total UL communication latency.

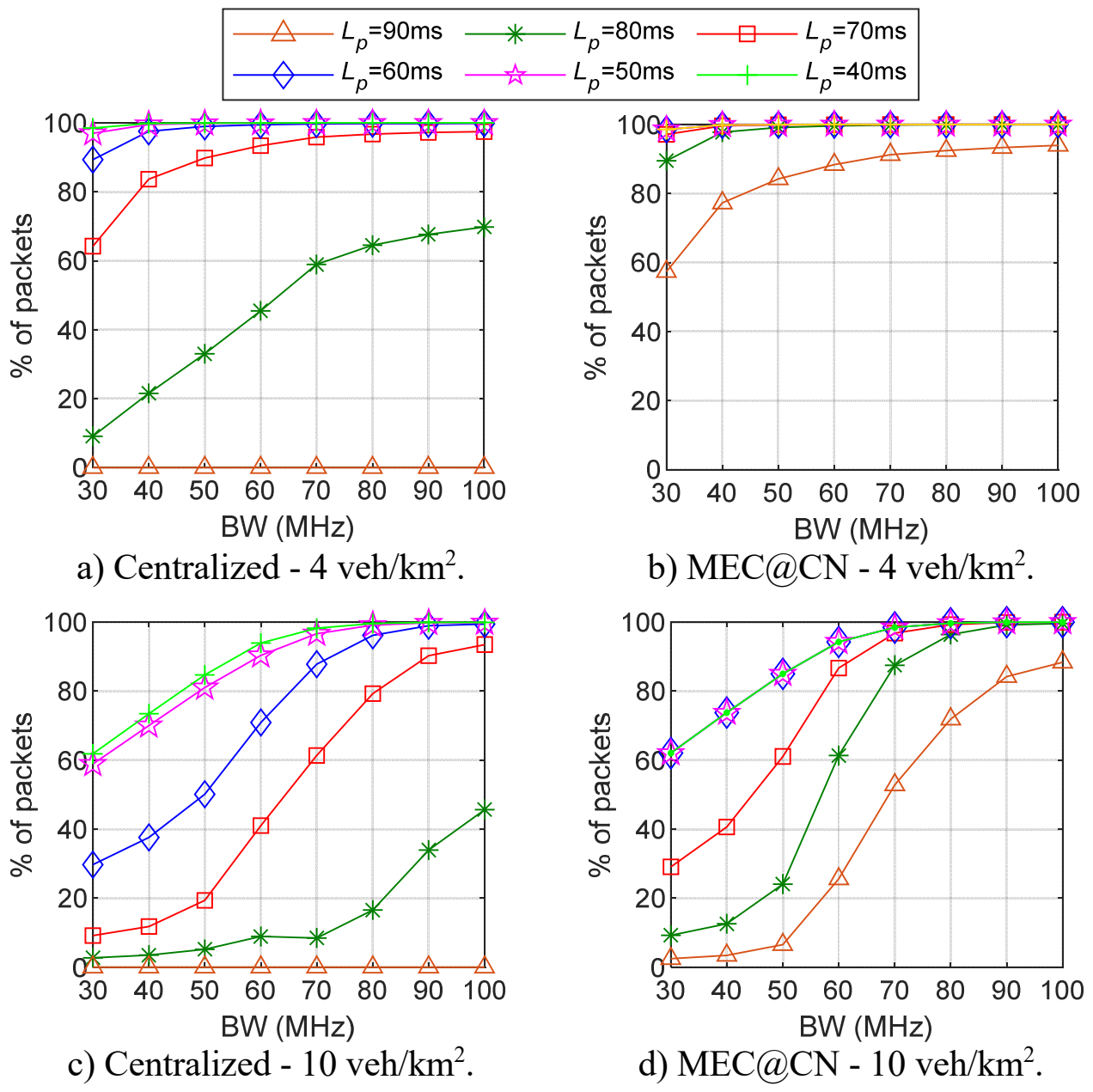

a) Centralized - 4 veh/km$^2$.
b) MEC@CN - 4 veh/km$^2$.
c) Centralized - 10 veh/km$^2$.
d) MEC@CN - 10 veh/km$^2$.

**Fig. 3**. Percentage of packets that meet the ToD UL latency requirement for different values of video processing delays (*Lp*).

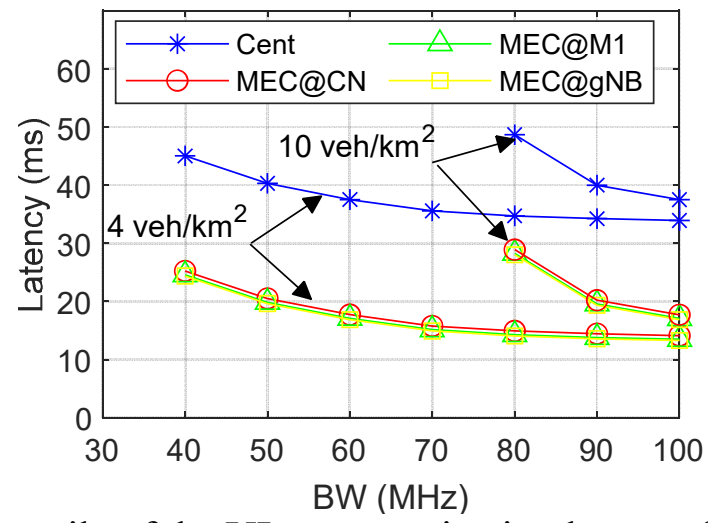

**Fig. 4**. 99$^{th}$ percentile of the UL communication latency for the centralized and MEC-based 5G network architectures using the TDD1 frame structure.

Fig. 3 also shows that the percentage of packets meeting the ToD latency requirement decreases as the density of vehicles increases. This decline is attributed to the increase in the UL communication latency (Fig. 4) due to higher latency values in the radio access network, as more vehicles compete for the available radio resources. The rise in radio latency with the vehicular density also leads to vehicles dropping some packets or video frames before they are transmitted. Specifically, vehicles drop a packet that is awaiting transmission if a newer packet with updated video images is generated. Fig. 3 and Fig. 4 show that the impact of higher vehicular densities affects all 5G network architectures. However, MEC-based architectures can still meet the ToD UL requirements with values of *Lp* up to 80 ms by increasing *BW*.

Table IV reports the bandwidth *BW* required to meet the ToD UL requirements for different video processing delays (*Lp*). The table indicates that MEC-based network architectures demand less bandwidth than a centralized network architecture to support the ToD UL requirements when *Lp* is high ($Lp \geq 60$ ms). Additionally, the table shows a significant increase in the required bandwidth with higher video processing delays *Lp*, with many scenarios unable to meet the ToD UL requirements even with 100 MHz (marked '-' in Table IV). This challenge is particularly evident for the centralized network architecture, which fails to satisfy the ToD UL requirements when *Lp* exceeds 60 ms. On the other hand, MEC-based architectures can meet the ToD requirements even when *Lp* is 80 ms with a bandwidth of 50 or 90 MHz for vehicle densities of 4 and 10 veh/km$^2$.

### B. Downlink communications

The ToD service requires the DL transmission of periodic command messages from the ToD control center to the vehicle. 5GAA establishes that these commands must be received in less than 20 ms with a reliability of 99.999%. Considering the 10 ms processing delay for the execution of car commands [10], the DL communication latency must be lower than 10 ms. Fig. 5 depicts the 99.999$^{th}$ percentile of the DL communication latency. Fig. 5 indicates that the centralized network architecture fails to meet the ToD DL requirements, as the 99.999$^{th}$ percentile of the DL communication latency exceeds ($\geq$54 ms) the maximum possible DL communication latency (10 ms) for all evaluated bandwidths and vehicle densities. The latency performance observed in the centralized network architecture is largely influenced by the latency of the Internet connection to the ToD control center. Even with a relaxation of the reliability requirement to 99%, the 99th percentile of the DL communication latency for the centralized network architecture consistently exceeds 22 ms (Cent-99tile in Fig. 5). On the other hand, Fig. 5 shows that MEC-based architectures can fulfill the ToD DL requirements, as the 99.999th percentile of the DL communication latency is below 10 ms for both vehicle densities and all evaluated cell bandwidths.

TABLE IV. BANDWIDTH (*BW* IN MHZ) NEEDED TO SATISFY THE TOD UL REQUIREMENTS

| **Vehicle density** | | **4 veh/km$^2$** | | | | | | **10 veh/km$^2$** | | | | | |
|---|---|---|---|---|---|---|---|---|---|---|---|---|---|
| ***Lp* (ms)** | | **40** | **50** | **60** | **70** | **80** | **90** | **40** | **50** | **60** | **70** | **80** | **90** |
| TDD1 | Cent. | 40 | 40 | 60 | - | - | - | 80 | 80 | 90 | - | - | - |
| | MEC@CN | 40 | 40 | 40 | 40 | 50 | - | 80 | 80 | 80 | 80 | 90 | - |
| TDD2 | Cent. | 40 | 50 | - | - | - | - | 80 | 90 | - | - | - | - |
| | MEC@CN | 40 | 40 | 40 | 50 | - | - | 80 | 80 | 80 | 90 | - | - |
| TDD3 | Cent. | 30 | 30 | 30 | - | - | - | 40 | 40 | 40 | - | - | - |
| | MEC@CN | 30 | 30 | 30 | 30 | 30 | 100 | 40 | 40 | 40 | 40 | 40 | - |
| FDD | Cent. | 30 | 30 | 30 | 60 | - | - | 40 | 40 | 40 | 80 | - | - |
| | MEC@CN | 30 | 30 | 30 | 30 | 30 | 60 | 40 | 40 | 40 | 40 | 60 | 80 |

## VI. Impact of the 5G Network Configuration

This section analyzes the impact of the 5G network configuration to support ToD. Specifically, we analyze the effects of the 5G duplexing mode and the configuration of the control channels on the 5G communication latency.

### *A. Duplexing mode*

Previous results were obtained using the TDD1 frame structure recommended by GSMA. Fig. 6 examines the impact of the duplexing mode and TDD frame structures on the 99th percentile of the UL communication latency for the centralized and MEC@CN architectures and different bandwidths. The results are presented for 4 veh/km$^2$, although similar trends were observed for 10 veh/km$^2$.

TDD1 (DDDDU) and TDD2 (DDDDDDDDUU) are the most commonly used TDD frame structures in 5G deployments [3]. Both TDD1 and TDD2 allocate 20% of the resources for UL transmissions. However, Fig. 6 shows that TDD2 increases the UL communication latency by approximately 10 ms compared to TDD1 under all evaluation conditions. This is attributed to the fact that TDD2 increases the time difference between consecutive UL slots, leading to a longer waiting time to access UL radio resources and thus delaying the transmission of data and control messages. Consequently, TDD2 requires more bandwidth than TDD1 to support the ToD UL requirements as shown in Table IV. For instance, TDD1 can meet the ToD UL requirements with a bandwidth of 40 MHz in a MEC@CN architecture with 4 veh/km$^2$ when $Lp$ is 70 ms. In contrast, TDD2 needs at least 50 MHz to fulfill the same requirements. Table IV also indicates that TDD2 is incapable of supporting the ToD UL requirements when $Lp > 80$ ms, even under the lowest density and with the MEC@CN architecture due to its higher UL communication latency (Fig. 6). TDD1 can meet the requirements under the same conditions with only 50 MHz. It is also worth noting that a more balanced allocation of radio resources between the UL and DL (FDD or TDD3) reduces the UL communication latency (Fig. 6) and the bandwidth needed to meet the ToD UL requirements (Table IV).

Finally, we should note that the duplexing mode does not have a significant impact on the DL communication latency. Similar latency values to those reported with TDD1 (Section V.B) were observed with TDD2, TDD3 and FDD. This is primarily due to the small network load generated by the ToD DL control messages in comparison to the quantity of radio resources allocated for DL transmissions.

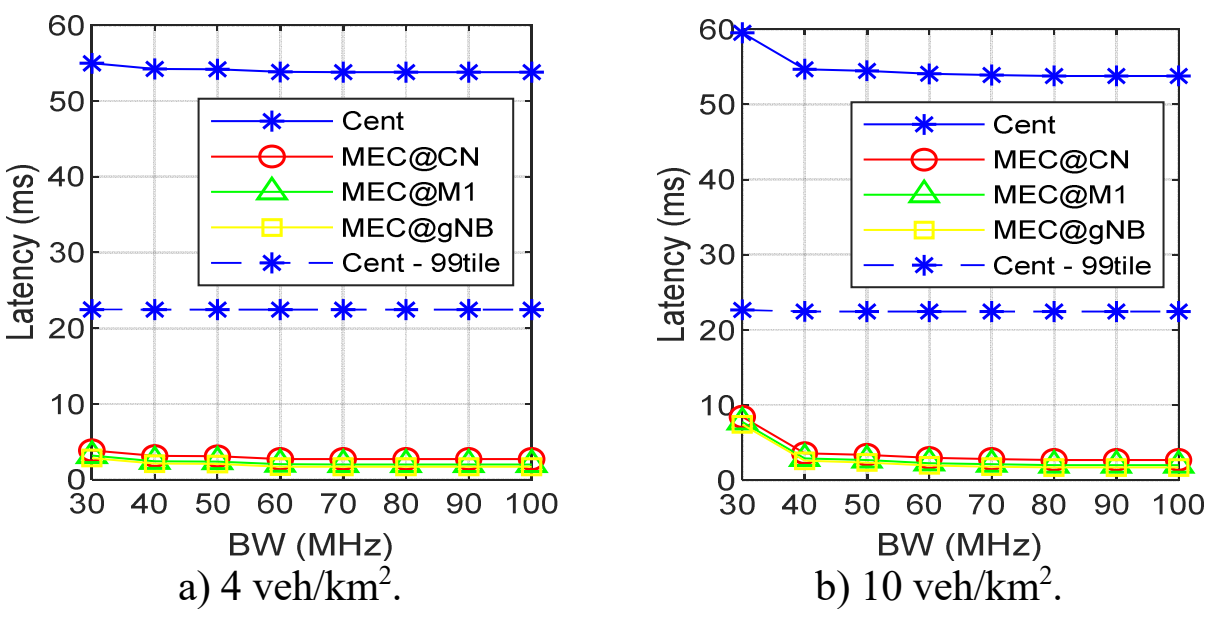

a) 4 veh/km$^2$. b) 10 veh/km$^2$.

**Fig. 5**. 99.999th percentile of the DL communication latency for centralized and MEC-based 5G network architectures using the TDD1 frame structure.

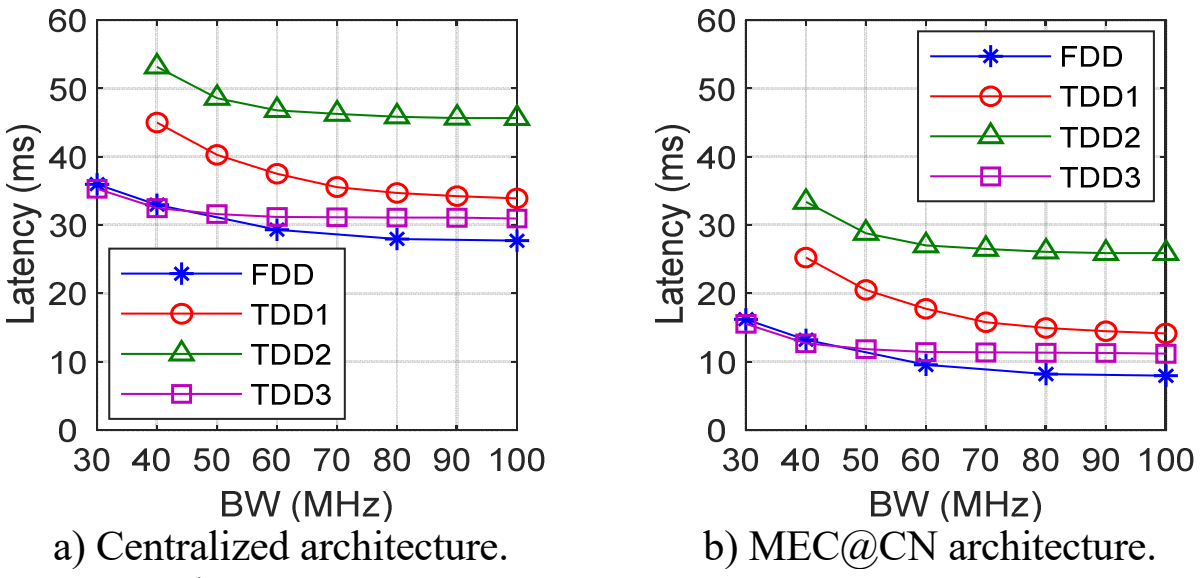

a) Centralized architecture. b) MEC@CN architecture.

**Fig. 6**. 99th percentile of the UL communication latency for different duplexing modes under a vehicle density of 4 veh/km$^2$.

### *B. Control channels*

Previous results have demonstrated that the TDD frame structure impacts the ToD UL communication latency because of the time difference between UL slots. This difference impacts the time vehicles must wait for UL data and control packets to be transmitted. These results were obtained using the SC1 slot configuration illustrated in Fig. 7 (for TDD1 and TDD2) [1]. With SC1, UL control messages are only transmitted in the last symbol of each UL slot, while DL control messages are transmitted in the first symbol of each DL slot. We also evaluate an alternative slot configuration for TDD1 and TDD2 (SC2 in Fig. 7) that is considered in 3GPP [17]. This configuration increases the opportunities for transmitting UL and DL control messages, which can reduce the communication latency. In the SC2 configuration, DL and UL control channel messages are transmitted in the first and last symbol of each slot, respectively (Fig. 7).

Fig. 8 depicts the percentage of packets that meet the 100 ms ToD UL latency requirement for various values of the video processing delays $Lp$ when employing the SC1 and SC2 slot configurations with the TDD1 and TDD2 frame structures. The figure shows that the utilization of SC2 increases the percentage of packets that meet the latency requirement, as SC2 decreases the UL communication latency (Fig. 9) by allowing UL control packets to be transmitted in all the slots of the frame. The improvements observed with SC2 compared to SC1 augment with increasing values of $L_p$ and the utilization of the TDD2 frame structure. This is because TDD2 increases the time difference between consecutive UL slots, and therefore benefits more from the ability to transmit UL control packets in all slots of the frame. Additionally, Table V demonstrates that the latency improvements achieved with SC2 reduce the bandwidth necessary to support the ToD service, particularly for higher values of $L_p$. For example, using SC2 with TDD1 can meet the ToD UL requirements when *Lp* is 90 ms using 100 MHz, while this is not possible with SC1. Similarly, the use of SC1 and TDD2 cannot meet the ToD UL requirements when *Lp* is higher than 70 ms, while SC2 and TDD2 can meet such requirements even when *Lp* increases to 85 ms.

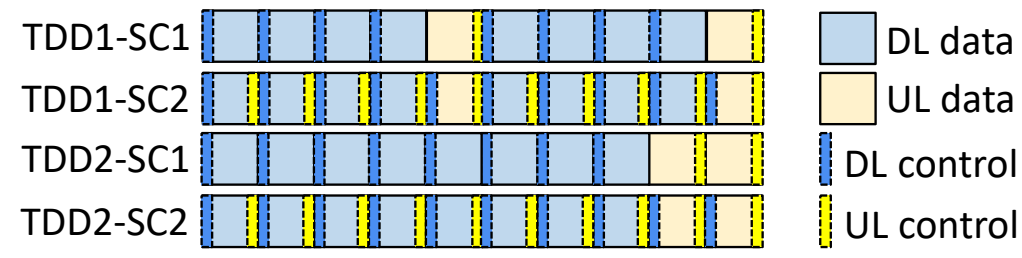

**Fig. 7**. TDD frame and slot configurations evaluated.

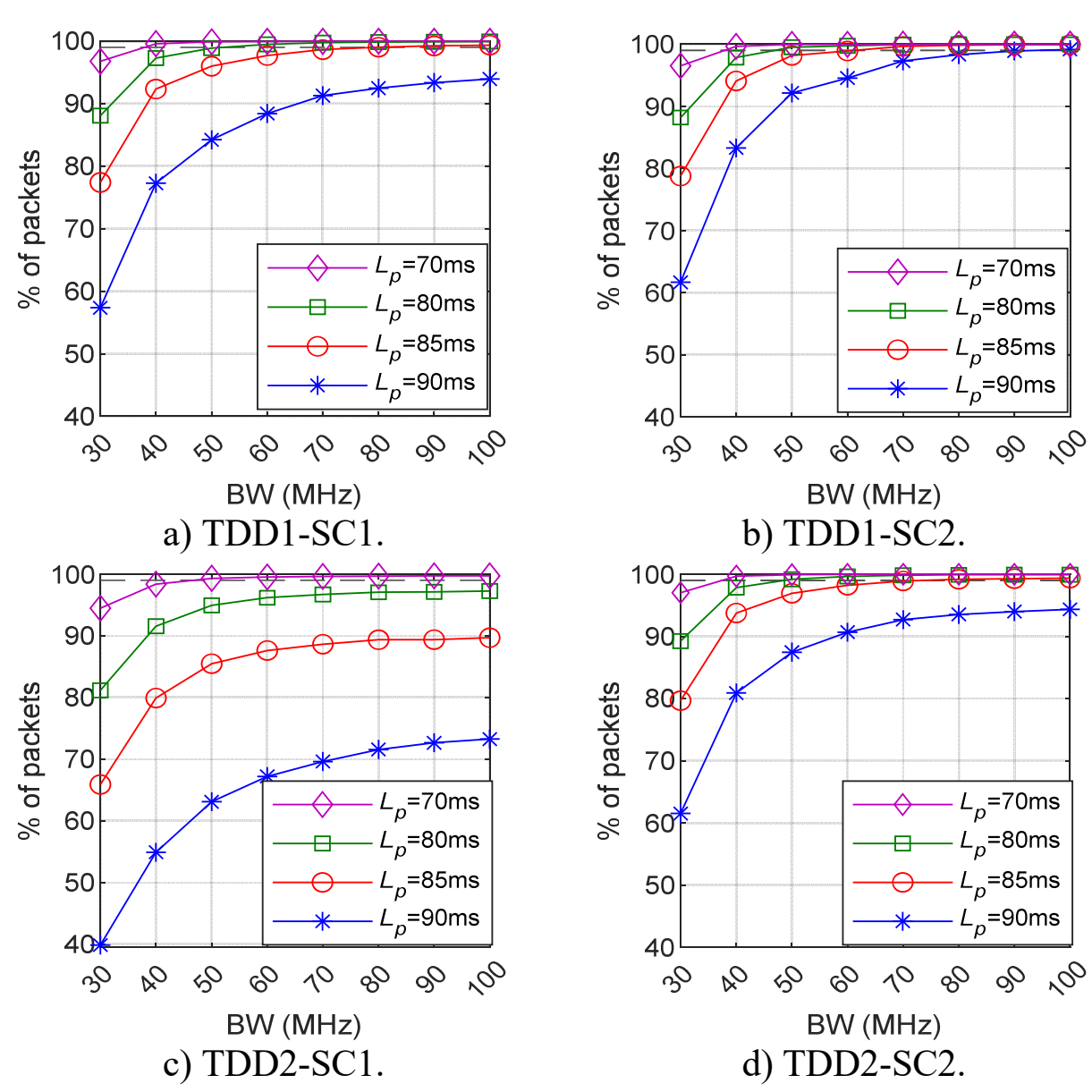

a) TDD1-SC1. b) TDD1-SC2. c) TDD2-SC1. d) TDD2-SC2.

**Fig. 8**. Percentage of packets that satisfy the ToD UL requirements for different values of $Lp$ (4 veh/km$^2$ and MEC@CN architecture).

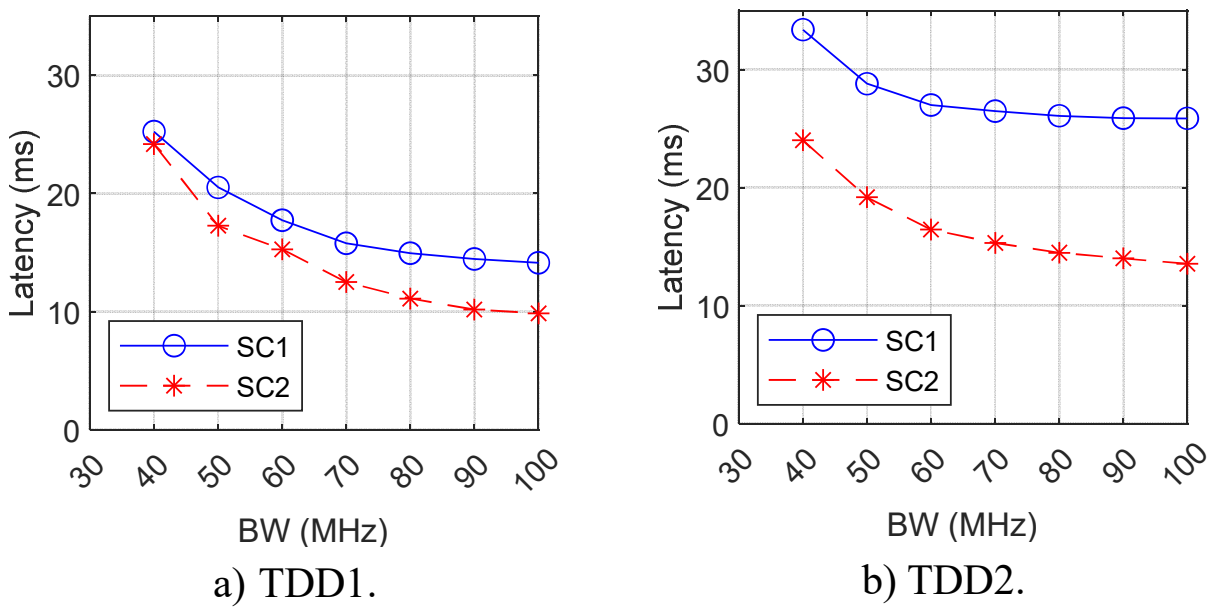

**Fig. 9**. 99th percentile of the UL communications for different TDD frame and slot configurations (4 veh/km$^2$ and MEC@CN architecture).

TABLE V. BANDWIDTH (*BW* IN MHZ) NEEDED TO SATISFY THE TOD REQUIREMENTS USING DIFFERENT SLOT CONFIGURATIONS (MEC@CN ARCHITECTURE)

| Vehicle density | | 4 veh/km$^2$ | | | | | | 10 veh/km$^2$ | | | | | |
|---|---|---|---|---|---|---|---|---|---|---|---|---|---|
| ***Lp*** **(ms)** | | 50 | 60 | 70 | 80 | 85 | 90 | 50 | 60 | 70 | 80 | 85 | 90 |
| TDD1 | SC1 | 40 | 40 | 40 | 50 | 80 | - | 80 | 80 | 80 | 90 | - | - |
| | SC2 | 40 | 40 | 40 | 50 | 70 | 100 | 80 | 80 | 80 | 90 | 100 | - |
| TDD2 | SC1 | 40 | 40 | 50 | - | - | - | 80 | 80 | 90 | - | - | - |
| | SC2 | 40 | 40 | 40 | 50 | 70 | - | 80 | 80 | 80 | 90 | - | - |

## VII. CONCLUSIONS

This study has analyzed the feasibility of 5G networks to support and scale ToD. The study has demonstrated that MEC-based network architectures are better suited to support ToD than centralized networks, which are constrained by the latency introduced by the Internet connection to the ToD control center. MEC-based architectures can better satisfy the UL and DL service requirements, and require less bandwidth to support and scale ToD. This is particularly the case with the most common TDD frame structures that allocate more slots for DL transmissions. The support of ToD is constrained by uplink video transmissions, and the capability to support the service depends on the video processing delays.

The study has demonstrated that the DDDSU TDD frame structure (TDD1) is better suited to support ToD than the DDDDDDDSUU TDD frame structure (TDD2), as it can better tolerate higher video processing delays and reduces the bandwidth necessary to scale the service. For instance, TDD1 only needs 40 MHz to support a ToD density of 4 veh/km$^2$ when the video processing delay is maintained lower than 70 ms, while TDD2 requires 50 MHz. The bandwidth required increases to 80 and 90 MHz for TDD1 and TDD2, respectively, when the ToD density is 10 veh/km$^2$ and the video processing delay is below 70 ms. The use of more UL/DL balanced TDD frame structures or FDD can support higher video processing delays and reduce the bandwidth necessary to scale the ToD service; for instance, they only need 40 MHz to support a ToD density of 10 veh/km$^2$. The study has also demonstrated that alternative configurations of the control channel, which increase the frequency of UL and DL control transmissions, can reduce the bandwidth necessary to scale ToD and better cope with higher video processing delays.

The study has demonstrated that MEC-based 5G network architectures are better suited to support the ToD service. The alternative network locations where the MEC can be deployed have shown only small differences in terms of communication latency. However, the decision on the MEC location should consider other important aspects like the number of ToD control centers required to support the ToD service in the road network, the MEC processing and computing power, and the signalling (and latency) required to guarantee service continuity as a ToD vehicle changes the serving MEC node.